\documentclass[aps,prl,twocolumn,noshowpacs,superscriptaddress,groupedaddress]{revtex4}
\usepackage{graphicx}  
\usepackage{dcolumn}   
\usepackage{bm}        
\usepackage{amssymb}   

\usepackage[T2A]{fontenc}

\usepackage[utf8]{inputenc}

\usepackage[english,russian]{babel}

\usepackage{amsmath}

\usepackage{amsfonts}

\usepackage{textcomp}

\usepackage{bm}

\usepackage{color}

\usepackage{verbatim}

\usepackage{subfig}

\begin{document}

\title{On consistency of hydrodynamic approximation for chiral media}

\author{A. Avdoshkin}
\affiliation{ITEP, B. Cheremushkinskaya 25, Moscow, 117218 Russia}
\affiliation{Moscow Inst Phys \& Technol, Dolgoprudny, 
Moscow Region, 141700 Russia.}

\author{V.P. Kirilin}
\affiliation{ITEP, B. Cheremushkinskaya 25, Moscow, 117218 Russia}
\affiliation{Department of Physics, Princeton University,
Princeton, NJ 08544, USA}

\author{A.V. Sadofyev}
\affiliation{ITEP, B. Cheremushkinskaya 25, Moscow, 117218 Russia}
\affiliation{Center for Theoretical Physics, 
Massachusetts Institute of Technology, Cambridge, MA,  02139}

\author{V.I. Zakharov}
\affiliation{ITEP, B. Cheremushkinskaya 25, Moscow, 117218 Russia}
\affiliation{Moscow Inst Phys \& Technol, Dolgoprudny, 
Moscow Region, 141700 Russia.}
\affiliation{School of Biomedicine, Far Eastern Federal University, 
Sukhanova str 8, Vladivostok, 690950, Russia}

\begin{abstract}

We consider chiral liquids, that is liquids consisting of massless fermions
and right-left asymmetric. In such media, one expects existence of 
 electromagnetic current flowing along an external magnetic field,
associated with the chiral anomaly. 
The current is predicted to be  dissipation-free.
We consider dynamics of chiral liquids, concentrating on the issues of possible 
instabilities and infrared sensitivity. Instabilities arise, generally speaking,
already in the limit of vanishing electromagnetic constant,
$\alpha_{el}\to~0$. In particular, liquids with non-vanishing chiral chemical
potential might decay into right-left asymmetric states containing vortices. 
\end{abstract}

\maketitle

\section{Introduction}

 Interest in theory of chiral liquids
was originally  boosted by the discovery of the quark-gluon plasma (since the light 
quarks are nearly massless).
 In theoretical studies,  one mostly concentrates, however, on a generic plasma
of massless fermions which interact in a chiral-invariant way 
and possess U(1) charges.
A remarkable feature of the chiral materials is  that the chiral anomaly,
 which is a loop, or quantum effect,
is predicted to have macroscopic consequences and effectively modifies 
the Maxwell equations.
In particular, in the equilibrium there is an electric current $j_{\mu}^{el}$
proportional to external magnetic field
\cite{vilenkin,chianov,kharzeev}:
\begin{equation}\label{cme}
j_{\mu}^{el}~=~\sigma_M B_{\mu}~~.
\end{equation}
Here $B_{\mu}$ is the magnetic field in the rest frame of the element of
the liquid, or $B_{\mu}\equiv~(1/2)\epsilon_{\mu\nu\alpha\beta}u^{\nu}F^{\alpha\beta}$,
where $u^{\mu}$ is the  4-velocity  of an element  of the liquid.

 In most recent times, the interest in theory of chiral liquids was triggered
by the paper in Ref. \cite{surowka} where it was demonstrated 
that the value of $\sigma_M$ is uniquely fixed  
in the hydrodynamic approximation. In particular,
for a single (massless) fermion of charge $e$:
\begin{equation}
\sigma_M~=~\frac{e^2\mu_5}{2\pi^2}~~,
\end{equation}
where $\mu_5$ is the chiral  chemical potential,
$\mu_5~\equiv~\mu_L-\mu_R$, so that $\mu_5\neq 0$ implies
 that the medium is not invariant under parity transformation.
  
Using the hydrodynamic approximation and equations of motion
one can demonstrate that
the magnetic conductivity is protected against corrections
\cite{surowka}. This non-renormalization of $\sigma_M$ goes back to
the Adler-Bardeen theorem.
Moreover, one can argue that the current
(\ref{cme}) is dissipation free \cite{yee}.  Indeed, both the r.h.s.
and l.h.s.
of (\ref{cme}) are odd under
time reversal.
This is a strong indication  that the dynamics behind
(\ref{cme}) is Hamiltonian and there is no dissipation \cite{yee}.
Analogy to
the superconducting current in the London limit,
$\vec{j}^{el}~=~m_{\gamma}^2\vec{A}$ where
$m_{\gamma}$ is the photon mass and $\vec{A}$ is the  vector potential,
supports \cite{zakharov2} the conclusion on the dissipation-free nature
of the current (\ref{cme}).

Further studies of dynamics of chiral liquids seem desirable. 
It is  worth emphasizing  that some chiral effects survive 
in the limit of the electromagnetic coupling
tending to zero, $\alpha_{el}\to~0$. In
 particular there is so called  chiral vortical effect
\cite{surowka}
which is the flow of axial current along the liquid's vorticity, $
j_{\alpha}^5~\sim~\epsilon_{\alpha\beta\gamma\delta}
u^{\beta}\partial^{\gamma}u^{\delta}$,
where $u_{\alpha}$ is the 4-velocity of an element of the liquid. 

In view of the
indications that chiral liquids possess such unusual properties.
In this note we will consider dynamics of the chiral liquids, concentrating mostly
on possible instabilities and infrared sensitivity. In particular, we 
argue that the instabilities arise already
in the limit $\alpha_{el}\to 0$.
Namely, chiral liquids with non-vanishing chemical potential
might decay
into a right-left asymmetric state containing vortices. 
The basic element of our analysis is consideration of 
consequences from conservation
of the axial current in the hydrodynamic approximation.  

Turn first to the definition of the axial charge 
on the fundamental, field-theoretic level:
\begin{equation}\label{operator}
Q^A~=~Q^A_{naive} + \frac{e^2}{4\pi^2}\mathcal{H}, ~~~\frac{d}{dt}Q^A~=~0~,
\end{equation}
where $Q^A_{naive}$ is the axial charge which is conserved according to
the classical equations of motion (without account of the anomaly) and
  $\mathcal{H}$ is the so called magnetic helicity:
\begin{equation}\label{helicity}
\mathcal{H}=\int \vec{A}\cdot \vec{B} d^3x~~,
\end{equation}
where $\vec{B}$ is the magnetic field and $\vec{A}$ is the corresponding 
vector potential.

In the approximation of external fields consideration of (\ref{operator})
would not bring any new insight compared to the approach of Ref. \cite{surowka}.
However Eq. (\ref{operator}) becomes more informative in case
of dynamical electromagnetic fields. The point is that, according to intuition
based on thermodynamics, all degrees of freedom contributing 
to the axial charge (\ref{operator}) are to be manifested in the equilibrium.
This implies that if one starts, for example, with a state $Q^A_{naive}~\neq~0,~
\mathcal{H}~=~0,$ then this state is in fact unstable and a
non-vanishing $\mathcal{H}$ would emerge spontaneously.
These expectations 
were verified in \cite{redlich,shaposhnikov,akamatsu,khaidukov} 
where the corresponding negative mode was identified
explicitly.  
 
Turn now to the hydrodynamic setup,  as it is   introduced in Ref.  \cite{surowka}.
Here, one assumes that there exists a liquid whose
constituents interact in a chiral invariant way. The liquid is described by the
standard hydrodynamic (relativistic) equations. To probe properties
of the liquid one introduces 
coupling with external electromagnetic fields. Mostly, 
the electromagnetic field is considered to be not dynamic.
Which means, in particular, that one neglects electromagnetic interactions between the
constituents as well as excitation of electromagnetic waves in the medium.
Within this framework one can send the the electromagnetic coupling
$\alpha_{el}$ to zero, $\alpha_{el}\to~0$ without affecting the properties of the
medium. The electromagnetic coupling survives only as a coefficient in front of, say, the
chiral magnetic current.  We will work within this framework when
electromagnetic interaction between constituents is 
absent or overshadowed by another, stronger 
interaction. Note that this framework differs from, say, magnetohydrodynamics.
In the latter case the electromagnetic field is dynamic and this is crucial to
establish electromagnetic instabilities 
\cite{redlich,akamatsu,shaposhnikov,khaidukov}.
We have mentioned  these results to introduce the issue of 
chiral-liquid instabilities.  Our central point, on the other hand,
 is that instabilities of chiral liquids exist also
in case of non-dynamic electromagnetic field, or even in case of neutral constituents.

In the hydrodynamic approximation, the axial current corresponding
to $Q^A_{naive}$ takes the form $j_{\mu}^A~=~n^Au_{\mu}$ where $n^A\equiv n^L-n^R$
and $n^{L,R}$ are the densities of the left- and right-handed
fermionic constituents, respectively. What is much less trivial, is that because of
the anomaly the axial current
contains further terms. 
Indeed, already the analysis of Ref. \cite{surowka} reveals
existence of the chiral vortical effect, with axial current being contributed 
by helical motion of the liquid (see also below). 

In more detail, the  axial charge in hydrodynamics can be represented as:
\begin{equation}\label{extended}
Q^A_{hydro}~=~Q^A_{naive}~+~Q^A_{mh}~+~Q^A_{mfh}~+~Q^A_{fh}~,
\end{equation}
where indices $``mh'' , ``fh''$ and ~$``mfh''$ stand for 
``magnetic helicity'' , ``fluid
helicity'' and mixed ``magnetic-fluid helicity'', respectively. 
Note that we are
using here the standard terminology  of magneto-hydrodynamics, 
see, e.g., \cite{bekenstein,moffatt},
where the fluid and magnetic helicities were considered phenomenologically,
without reference to chiral liquids.
In particular,
 $Q^A_{mh}$ stands for $(e^2/2\pi^2)\mathcal{H}$.  
The fluid helicity, $Q^A_{fh}$ is defined as the charge associated with
the  current  $j^{\alpha}_{fh}$: 
\begin{equation}\label{zero}
Q^A_{fh}~=~\frac{1}{4\pi^2}\int d^3 xj^0_{fh}~~,
\end{equation} 
while the current $j^{\alpha}_{fh}$ is given by:  
\begin{equation}\label{fluidcurrent}
j^{\alpha}_{fh}~=~2\mu^2 \omega^{\alpha}~~,
\end{equation}
where 
\begin{equation}\label{omega}
\omega^{\alpha}~\equiv~\frac{1}{2}\epsilon^{\alpha\beta\sigma\rho}u_{\beta}\partial_{\sigma}u_{\rho}.
\end{equation}
An explicit expression for the mixed helicity is given later. Actually, 
the algorithm of construction
of various pieces  in (\ref{extended}) is readily identified 
by using analogy with the pure
electromagnetic case, as we explain in a moment.

Eq. (\ref{extended}) can be substantiated in a number of ways. One possibility
is to look into explicit expressions for vector and axial currents obtained via the
procedure introduced first in \cite{surowka}. We will come to this point later,
see discussion of Eq. (\ref{close}) below.
 Now, we will follow \cite{shevchenko} to argue that in the
hydrodynamic setup one should substitute the standard 
electromagnetic potential $eA_{\mu}$ in the expression 
for the chiral anomaly by the following combination:
\begin{equation}\label{shevchenko}
eA_{\mu}~\to~eA_{\mu}~+~\mu u_{\mu} ~,
\end{equation}
where $\mu$ is the chemical potential associated with the charge, source of the
potential $A_{\mu}$.

The most straightforward  way to justify (\ref{shevchenko}) is to observe
that chemical potential thermodynamically is introduced through an
extension of the original Hamiltonian $\hat{H}$:
\begin{equation}\label{hydro}
\hat{H}~\to~\hat{H}~-~\mu\hat{Q}~-~\mu_5\hat{Q}^A~.
\end{equation}
As far as the chemical potentials  $\mu,\mu_5$ are  considered to be small,
the corresponding change in the Lagrangian  $\delta L$ is given by
\begin{equation}\label{deltaL}
\delta L=-\delta H~=~\mu Q +\mu_5Q^A=~\mu\bar{\psi}\gamma_0\psi+
\mu_5\bar{\psi}\gamma_0\gamma_5\psi .
\end{equation}
The next step is to generalize (\ref{deltaL}) to the case of hydrodynamics.
The generalization assumes rewriting (\ref{deltaL}) in an explicitly
Lorentz-covariant way:
\begin{equation}
\delta L~=~\mu u^{\alpha}\bar{\psi}\gamma_{\alpha}\psi+
\mu_5u^{\alpha}\bar{\psi}\gamma_{\alpha}\gamma_5\psi~ .
\end{equation}
In case of $\mu_5=0$, 
the substitution (\ref{shevchenko}) becomes then obvious.

As a result of substitution (\ref{shevchenko})
the definition of the conserved axial charge (\ref{operator}) 
is generalized in
hydrodynamics  to the expression (\ref{extended}), and we will
consider implications of this extension
\footnote{Note, that generally speaking
there is also a 
thermal contribution to $Q_{fh}$ 
which is not captured by the substitution (\ref{shevchenko}). 
It can be derived by considering field theory 
in a non-inertial reference frame \cite{vil}. 
In this note we omit this contribution for simplicity 
and in what follows separate it from the "fh" term.}.

 Consider  a non-vanishing chiral chemical potential, 
$\mu_5\neq 0$.
Then one expects that in the equilibrium all 
the degrees of freedom with a non-vanishing
axial charge are  excited and all the helicities entering (\ref{extended})
are non-vanishing. This implies, in turn, that if 
one starts with the state where the whole
axial charge is attributed to a single term 
in the r.h.s. of (\ref{extended}), say, to
the charge of elementary constituents,
$$Q^A_{naive}~\neq~0,~~Q^A_{mh}~=~Q^A_{fh}~=~Q^A_{mfh}~=0~,$$
then this state is unstable with respect 
to generation of all types of helicities.
The instability with respect 
to generation of  magnetic fields with
non-vanishing helicity (\ref{helicity}) 
was considered in detail and in various applications
\cite{redlich,shaposhnikov,akamatsu,khaidukov}. Eq.
(\ref{extended}) implies that in fact
one can expect that in hydrodynamics  there are
more general instabilities which would result in generation
of all possible helicities:
\begin{equation}\label{equal}
Q^A_{naive}~\sim~Q^A_{mh}~\sim~Q^A_{mfh}~\sim~Q^A_{fh}~.
\end{equation}
In other words, all types of helical motions are excited
in  chiral plasma on  macroscopic scales.

 It is amusing that the possibility of the helicity conservation
in the ordinary hydrodynamics
(without any reference to the chiral liquids) has been studied in great detail,
for review see, e.g., \cite{bekenstein,moffatt}. The generic conclusion is that
the fluid helicities are  conserved in the limit of vanishing dissipation.

The outline of the paper is as follows. In the next section we address the issue of
infrared sensitivity of the definition of axial charge in field theory. Next, we 
substantiate representation of the axial charge in hydrodynamics as a sum of various
vorticities. Then, we argue that the magnetic and fluid helicities are conserved in 
classical limit in case of absence of dissipation. Next, we
 introduced various types of instabilities of chiral liquids. 

In conclusion, we summarize the results obtained.

\section{Evaluation of axial charge }

In this section we outline evaluation of 
the anomalous piece in the conserved axial charge
(\ref{operator}). In particular, we emphasize that the calculation
is actually valid only in the limit of exact symmetry. In other words, 
fermion masses are assigned to be exact
zero. Considering this limit is common to the recent papers on the
anomalous hydrodynamics, see, e.g. , \cite{jensen}. 
Only in this limit the effect
of the anomaly is reduced to local terms in the effective action.  Moreover,
there is no explicit time dependence as if we are discussing 
{\it static} processes.
A specific feature of such local, or polynomial
terms is that the action is gauge invariant while the density 
of the action is not gauge
invariant. The expression (\ref{helicity}) 
for the magnetic helicity provides the best known
example of such a term.

If one introduces explicit violation of the chiral symmetry, 
say, through the masses of
the constituents, then the
effect of the anomaly does not reduce to local terms in the
effective action.
 Nevertheless,
in a certain kinematic limit the matrix element of the axial current
 becomes again  the same polynomial
as in (\ref{operator}).
 We emphasize that this kinematic limit actually assumes non-vanishing
{\it time-dependent} fields. In particular, 
the expression (\ref{operator}) for the matrix element
of the
axial charge in the limit of electric fields much stronger 
than the fermionic masses:
\begin{equation}\label{limit}
m_f^2~\ll~E~\ll~H~~,
\end{equation}
where $E,H$ are electric and magnetic fields, respectively. 
The constraint (\ref{limit})
is mentioned in \cite{zakharov2}. 
Here we present a more detailed derivation of (\ref{limit}).

Thus, our aim here is to evaluate the matrix element of
the axial charge over a photonic state
$\langle\gamma|Q^A|\gamma\rangle$, where
\begin{equation}\label{axialcharge}
Q^A~=~\int d^3x j^A_0(\vec{x},t)~=~
\int d^3x \bar{\psi}\gamma_0\gamma_5\psi
\end{equation}
and $\psi$ is a massless Dirac field of
charge $e$. Moreover, consider temperature-zero case 
and the photons on  mass shell.
Then, it is well known that the matrix element of the axial current $j_{\mu}^A$
corresponding to the anomalous triangle graph
has a pole. In the momentum space,
\begin{equation}\label{current}
\langle \gamma|j_{\mu}^A|\gamma\rangle~=~\frac{e^2}{2\pi^2}\frac{iq_{\mu}}{q^2}
\epsilon_{\rho\sigma\alpha\beta}e^{(1)}_{\rho}k^{(1)}_{\sigma}e^{(2)}_{\alpha}k^{(2)}_{\beta}~~,
\end{equation}
where $q_{\mu}$ is the 4-momentum brought in by the axial current,
$e_{\rho}^{(1)},k^{(1)}_{\sigma}$ and $e^{(2)}_{\alpha}k^{(2)}_{\beta}$
are the polarization vectors
and momenta of the photons.

The matrix element (\ref{current}) of the axial current
 is clearly non-local in nature, by virtue
of the Lorentz covariance and gauge invariance. Concentrate, however, on the
matrix element of the axial charge
(\ref{axialcharge}).
Since the charge is defined as $Q^A=\int d^3x j_0^A(\vec{x},t)$,
evaluating the charge implies considering
the kinematical limit
$$\vec{q}~\equiv~0,~q_{0}~\to~0 ~~.$$
In this limit the matrix element (\ref{current}) reduces to a polynomial:
\begin{equation}
\langle\gamma|Q^A|\gamma\rangle~=i\frac{e^2}{4\pi^2} \epsilon^{ijk}
e^{(1)}_ie^{(2)}_j (k^{(1)}-k^{(2)})_k~
\end{equation}
and we come, indeed,  to the standard expression for the
magnetic helicity (\ref{operator}).

 Evaluating the charge (\ref{axialcharge}) starting from the non-local expression
(\ref{current})
for the current has  advantages, from the theoretical point of view. In particular,
we avoid considering contribution of heavy regulator fields, and our derivation
of (\ref{axialcharge}) is given entirely in terms of physical, or light (massless)
 degrees of freedom. On the other hand, the now-standard way of evaluating
the magnetic conductivity $\sigma_M$ is to relate it to the spatial correlator
of two electromagnetic currents
(for review see, e.g., \cite{landsteiner}). In the momentum space:
\begin{equation}\label{correlator}
\sigma_M~=~\lim_{q_0\equiv 0,{q}_k~\to 0}\epsilon^{ijk}\frac{i}{2q_k}
\langle j^{el}_i,~j^{el}_j\rangle
\end{equation}
Although taking the limit of $q_k\to 0$ 
implies, at first sight, that the correlator
(\ref{correlator}) is sensitive to large distances, 
$r~\sim~1/|\vec{q}|$, in fact,
it depends on the correct definition of the correlator at the coinciding points.
Therefore, one has to consider carefully 
the ultraviolet regularization procedure,
for details see \cite{landsteiner}.

Necessity of a careful treatment of the {\it time-dependent} fields looks
counter-intuitive in view of the fact that (\ref{correlator})
relates the magnetic conductivity to a pure spatial correlator. It might, therefore, worth
 reminding the reader that in the  original derivation of the 
axial anomaly in terms of zero modes
in magnetic field \cite{ninomiya} one evaluates actually the work $W$ produced by
an external {\it electric} field $\vec{E}$:
\begin{equation}\label{work}
W~\equiv	~\vec{E}\cdot\vec{j}_{el}~=~\vec{E}\cdot \vec{B}\frac{e^2}{2\pi^2}\mu_5~~.
\end{equation}
This work compensates the energy needed for massless pair production.
And only after cancelling the electric field from the both sides
of (\ref{work}) one arrives at the current
(\ref{cme}) which apparently depends on the magnetic field alone.

If, on the other hand, one
introduces finite fermionic masses 
then there is no pair production for
$E\ll m_f^2$ and the role of the time-dependent electromagnetic 
potentials is made explicit.
In particular, taking the limit $q_0~\to~0~ (\vec{q}~\equiv~0)$ now gives
$$\langle \gamma|Q^A|\gamma \rangle_{m_f\neq 0}~=~0~~,$$
since there is no singularity at $q^2=0$ in the matrix 
element corresponding to the triangle graph.

\section{Axial charge in hydrodynamics}

Probably, the most striking novel feature brought in by
consideration of  hydrodynamics is the
emergence of the  chiral vortical effect,  see, e.g.  \cite{surowka,review}, or
 the flow of the axial current along the fluid vorticity:
\begin{equation}\label{vortical}
j_{\alpha}^A~= ~ \sigma_{\omega}
\omega_{\alpha}~,
\end{equation}
with the vortical conductivity $\sigma_{\omega}~\neq ~0$.
The substitution (\ref{shevchenko}), in case of both $\mu,\mu_5~\ne ~0$,
fixes $\sigma_{\omega}$ as:
\begin{equation}\label{notemperature}
\sigma_{\omega}~=\frac{(\mu^2+\mu_5^2)}{2\pi^2}~~.
\end{equation} 
The reservation in using the substitution (\ref{shevchenko}) is that
it does not capture temperature dependences. Also,
in Eq. (\ref{notemperature}) we did not account for possible
spatial variations of the chemical potentials.

Although the chiral vortical effect  is rooted in the anomaly it
 survives
in the limit of vanishing electromagnetic coupling.
If we restore the term proportional to  $\sqrt{\alpha_{el}}$ then the
 axial-vector current looks as follows:
 \begin{eqnarray}\label{close}
j^A_{\mu}=n^Au_{\mu}+
\left(\frac{\mu^2+\mu_5^2}{2\pi^2}\right)\omega_{\mu}+
\frac{e\cdot \mu}{2\pi^2}B_{\mu}+O(e^2)\label{Acurrent}
\end{eqnarray}
where $n^A$ is the density of constituents and $B_{\mu}$ is defined in Eq. (\ref{cme}). 
The first term in the r.h.s. of Eq. (close) corresponds to
$Q^A_{naive}$ in Eq. (\ref{extended}), the second term corresponds to
$Q^A_{fh}$ and 
the third term in the r.h.s. of this equation corresponds to the 
so called mixed magneto-fluid helicity, see, e.g., \cite{bekenstein}.
The corresponding axial charge, $Q^A_{mfh}$ entering 
(\ref{extended}) is proportional to  the volume
integral of the temporal component of $j^{\alpha}_{mfh}$: 
\begin{equation}\label{one}
j_{mfh}^{\alpha}~=~
\frac{1}{2} \epsilon^{\alpha\beta\rho\sigma}A_{\beta}\omega_{\rho\sigma}
\end{equation}
where $\omega_{\alpha\beta} = \partial_{\alpha}(\mu u_{\beta})-\partial_{\beta}(\mu u_{\alpha})$.
There is also an alternative form of $j^{\alpha}_{mfh}$ defined as:
\begin{equation}\label{two}
j^{\alpha}_{mfh}~=~\frac{1}{2}  
\epsilon^{\alpha\beta\rho\sigma}(\mu u_{\beta})F_{\rho\sigma}~~.
\end{equation}
The corresponding charge, $Q^A_{mfh}$ is the same in 
the both cases of (\ref{one}) and (\ref{two}).

Eq. (\ref{close}) is close to the expression found first in the pioneering paper
\cite{surowka}, see also \cite{oz}. And, as is mentioned in the Introduction,
the expression for the current obtained in \cite{surowka}
could be considered as a motivation to introduce (\ref{extended}).
The main difference is that the coefficient in front of the fluid
vorticity $\omega_{\mu}$ in Ref. \cite{surowka} contains also terms of third power
in chemical potentials. Moreover, using the hydrodynamic expansion
to higher orders in derivatives, in the spirit of
the approach of \cite{surowka}
would bring further corrections to the current $j_{\mu}^A$.
The possibility of variations in explicit expressions for the current
is rooted in freedom of choosing
frames, or precise definition of the fluid velocity $u_{\mu}$. Ref. \cite{surowka}
uses the Landau frame introduced, e.g., in the
textbook of Landau \& Lifshitz,  while Eq. (\ref{close}) assumes the use of the so called
entropy frame, see, in particular, \cite{nicolis}.  

A crucial point is that the expressions for 
$Q^A_{naive}, Q^A_{fh}, Q^A_{mfh}, Q^A_{mh}$
do not receive further contributions in the expansion in hydrodynamic derivatives. 
The absence of higher
order terms in
the hydrodynamic expression  (\ref{close}) is a reflection of the important property
of the chiral anomaly on the fundamental level
that it is limited to a single term $\mathcal{H}$. 

So far we exploited   the substitution
(\ref{shevchenko}) following the argumentation of \cite{shevchenko}
reproduced above. However, this argumentation by itself 
is of somewhat heuristic nature
and it is worth emphasizing that similar conclusions were reached more recently
in a systematic way within the geometric approach to hydrodynamics,
see, e.g., \cite{megias,jensen}.

To derive the chiral 
effects in hydrodynamics within these approaches
one considers motion in both electromagnetic and gravitational backgrounds.
This seems to be rooted in the very nature of the hydrodynamics which
is entirely determined by
conservation laws, of energy-momentum tensor and of relevant currents.

Technically, one way to trace this kind of 
unification of electromagnetic
and gravitational interactions is to start with a covariant action
in higher dimensions. One can demonstrate then \cite {megias}  that the mixed
gauge-gravity anomaly in higher dimensions generates a
4d action which is responsible for the chiral effects.
 
Moreover  the conductivity $\sigma_{\omega}$ gets related \cite{megias}
to the correlator of components of the 
energy-momentum tensor and electric current:
\begin{equation}\label{sigmaomega}
\sigma_{\omega}~=~\lim_{q_0\equiv 0,q_k\to 0}\frac{i}{q_k}\epsilon^{ijk}\langle j^A_i,T_{0j}\rangle~.
\end{equation}
The connection of our procedure to that of \cite{megias}
can readily be established. 
Indeed, modification of the naively conserved
axial charge $Q^A_{naive}$ by the anomaly is in one-to-one correspondence with
the non-vanishing correlator (\ref{correlator}). 
The hydrodynamic modification (\ref{hydro}) of the
field theoretic Hamiltonian implies modification of the $T^{0i}$ component
of the energy-momentum tensor. Choosing, for simplicity, $\mu_5=0$,
$$(\delta T^{0i})_{hydro}~=~\mu J^i~~,$$
where $J^i~(i=1,2,3)$ are the spatial
components of the  vector current.
Therefore, there arises an anomalous piece in the correlator 
\begin{equation}\label{relation}
\epsilon_{ijk}\frac{\langle T^{0i},J^{j}\rangle}{q^k}~=~\mu\epsilon_{ijk}\frac{\langle
J^{i},J^{j}\rangle}{q^k}~.
\end{equation}
Eqs (\ref{correlator}),(\ref{sigmaomega}), (\ref{relation})
imply that conductivities $\sigma_M,\sigma_{\omega}$ are related
to each other and, actually, in exactly the same way as prescribed
by the substitution (\ref{shevchenko}).  
Following this logic, we derive the $Q^A_{fh}$ contribution to the axial 
charge in the hydrodynamic approximation (\ref{extended}).

Another geometric approach \cite{jensen}
starts with considering a static metric
$$ds^2~=~-\exp{ (2\sigma (\vec{x}))}
\big(dt+a_i(\vec{x})dx^i\big)^2+g_{ij}(\vec{x})dx^idx^j$$
There is also electromagnetic background ${A}_{\mu}(\vec{x})$. 
Then one can demonstrate
 that the symmetries of the problem imply that the partition function depends
 in fact on the combinations $\mathcal{A}_0$, $\mathcal{A}_i$,
\begin{equation}\label{KK}
e{A}_0~=~e\mathcal{A}_0~+~\mu~,~~e{A}_i~=~e\mathcal{A}_i~-~A_0a_i\end{equation}
which are Kaluza-Klein gauge invariant
and are replacing $e{A}_{\mu}$ in the standard field theoretic expressions.


Moreover, it is quite obvious that the procedure we are using has much in
common with the approach \cite{jensen}. 
Indeed, the field $a_i$ entering (\ref{KK})
is also proportional to $u_i$ and we readily come to the 
same contributions to the hydrodynamic axial charge as derived above. Note, however, that the approach
of \cite{jensen} applies only in equilibrium while the substitution (\ref{shevchenko}) works
in general case. 

This concludes the derivation of 
the axial charge  in the
hydrodynamic limit, see Eq. (\ref{extended}).
It is worth mentioning again that all the conclusions concerning 
chiral plasmas are subject
to the reservation that, from the microscopical point of view, 
the underlying field theories are assumed to be infrared stable.
Note  that within the
holographic approach it turns possible in some cases to study 
dynamics of chiral liquids in
infrared. There are indications that the physics 
in the infrared could be richer
than is usually assumed. In particular, new scales 
can be generated, for a recent
study and further references see \cite{marolf}.


\section{Classical conservation of magnetic and fluid helicities}
As it is mentioned above, possibility of conservation of the magnetic and fluid helicities
was intensely discussed in the context of the magnetohydrodynamics.
Here we will reproduce the main results in a way close to Ref.
\cite{bekenstein} 
Let us begin with the conservation of the fluid helicity.
The main  tool to be used is the 
relativistic 
version of the Euler equation: 
\begin{eqnarray}
(p + \epsilon)a^{\mu} = (-\partial^{\mu}p-u^{\mu}u^{\nu}\partial_{\nu}p) = 
-P^{\mu \nu}\partial_{\nu}p.
\end{eqnarray}
where $\epsilon$ and $p$ stand for proper energy density and pressure respectively,
$u^{\beta}$ is the 4-velocity normalized as $u_{\beta}u^{\beta}~=
~-1$, $a^{\mu} = u^{\nu}\partial_{\nu}u^{\mu}$ is the acceleration and 
$P^{\mu\nu}=u^{\mu}u^{\nu}+g^{\mu\nu}$ is the projection operator. 
Here electric field is switched off what will be justified later. Moreover, we will utilize
Gibbs-Duhem relation:
\begin{eqnarray}
d \epsilon = \rho d \mu + s dT.
\end{eqnarray}
where $\mu$ is
the chemical potential conjugated to the charge $\rho$, $T$
 is the local temperature and $s$ is proper density of entropy.

Now we investigate the behaviour of different parts of the axial current in a 
chiral-neutral charged fluid.
After some algebra, one can demonstrate that the current
$j^{\alpha}_{fh}$ associated with the fluid helicity has the following divergence:
\begin{eqnarray}\label{divergence0}
\partial_{\alpha} j^{\alpha}_{fh}
= \frac{2T^2\mu s}{p+\epsilon}\omega^{\alpha} \partial_{\alpha}\left( \frac{\mu}{T} \right).
\end{eqnarray}
where $j^{\alpha}_{fh}$ and $\omega^{\alpha}$ are defined in (\ref{fluidcurrent}) and
(\ref{omega}), respectively. Thus, if $\displaystyle \frac{\mu}{T}= 
const$ (e.g. $T\to0$) this contribution to the axial charge (\ref{zero}) 
is conserved individually.

Turn now to the mixed magnetic-fluid helicity in the absence of electric field. 
The divergence of the corresponding current (\ref{two}) is given by:
\begin{equation}\label{divergence1}
(j^{\alpha}_{fmh})_{,\alpha}~=~1/4\epsilon^{\alpha\beta\gamma\delta}
\omega_{\alpha\beta}F_{\gamma\delta}~.
\end{equation}
The next step is to express $F_{\alpha\beta}$ in terms of
$B_{\mu}$ when $E_{\nu}=0$ \cite{yodzis} :
\begin{equation}
F_{\alpha\beta}~=~\epsilon_{\alpha\beta\gamma\delta}B^{\gamma}u^{\delta}~.
\end{equation}
Using this as an input one comes to:
\begin{equation}
(j^{\alpha}_{fmh})_{,\alpha}~=~\frac{T^2\mu s}{p+
\epsilon}B^{\alpha} \partial_{\alpha}\left( \frac{\mu}{T} \right)~~,
\end{equation}
and the current is again conserved if $T\to0$ or $\displaystyle \frac{\mu}{T}= const$.

Finally, consider the magnetic helicity (\ref{helicity}). The corresponding 4d
current is defined as
\begin{equation}
j^{\alpha}_{mh}~=~
\frac{1}{2}\epsilon^{\alpha\beta\gamma\delta}A_{\beta}F_{\gamma\delta}~~.
\end{equation}
The divergence of this current is proportional to the product of magnetic and electric fields
$B_{\mu}$ and $E_{\mu}$,
\begin{equation}\label{divergence}
(j^{\alpha}_{mh})_{,\alpha}~=~-2B^{\mu}E_{\mu}~,\end{equation}
where $B_{\mu}$ is defined in (\ref{cme}),  $E_{\mu}~=~F_{\mu\nu}u^{\nu}$ 
and one finds that $Q_{mh}=\frac{e^2 }{4\pi^2}\mathcal{H}$ .

Eq. (\ref{divergence}) is pure kinematic in nature. The dynamic
input that
ensures conservation of the current $j^{\alpha}_{mh}$ 
is that in the case of zero-temperature perfect magneto-hydrodynamics 
$E^{\mu}$ is to vanish along with the temperature:
\begin{equation}
E_{\mu}~\to~0,~~if~~\sigma_E~\to~\infty~.
\end{equation}
One can also easily evaluate the dissipation rate of the magnetic helicity in 
the case of zero temperature and finite conductivity $\sigma_E$ 
in classical hydrodynamics \cite{moffatt}:
\begin{equation}
\frac{d\mathcal{H}}{dt}~=~\frac{-2}{\sigma_E}
\int d^3x~\vec{B}\cdot\mathbf{curl}~\vec{B}~~,
\end{equation}
for details of the derivation see, e.g., \cite{bekenstein}.

In the case of arbitrary temperature one can show that the
 divergence of the sum of the three helicity currents is
\begin{eqnarray}
2\partial_{\alpha}(\mu B^{\alpha}+\mu^2\omega^{\alpha})+
\frac{1}{4}\epsilon^{\alpha\beta\delta\gamma}F_{\alpha\beta}F_{\gamma\delta}=\nonumber\\
=-(F^{\alpha\beta}+\omega^{\alpha\beta})u_{\beta}(\tilde{F}_{\alpha\gamma}
+\tilde{\omega}_{\alpha\gamma})u^{\gamma}=\\
= -\frac{sT}{(\epsilon +p)}\left(E^{\alpha}-
TP^{\alpha\beta}\partial_{\beta}\left(\frac{\mu}{T}\right)\right)(\tilde{F}_{\alpha\gamma}
+\tilde{\omega}_{\alpha\gamma})u^{\gamma} \nonumber
\end{eqnarray}
and the similar result holds for the thermal part of the flow helicity:

\begin{eqnarray}
\partial_{\mu}(T^2 \omega^{\mu}) =\frac{2 T^2\rho}{\epsilon + p}
\omega^{\mu}\left(E_{\mu}-T\partial_{\mu}\left(\frac{\mu}{T}  \right)\right).
\end{eqnarray}

Thus charge conservation constraint is fulfilled in 
dissipationless limit when the conductivity 
related quantity $E^{\alpha}-TP^{\alpha\beta}
\partial_{\beta}\left(\frac{\mu}{T}\right)$ 
is zero. Which, in turn, corresponds to $\sigma \to \infty $. 
It is worth emphasizing that the viscosity also causes 
helicity to dissipate so as it was above for an ideal 
liquid there is an extra requirement $\eta \to 0$ for 
the axial current conservation to hold.

Another feature which unifies various types of helicity is that
the corresponding charges are related to linkage of magnetic and fluid vortices.
In particular, the fluid helicity is a measure of linkage of vortex lines
in the liquid \cite{moffatt3}. The fluid-magnetic helicity 
measures the linkage number of
closed vortex lines and magnetic flux lines. Finally, 
the magnetic helicity can be interpreted
in terms of the fluxes of linked flux tubes.

This relation of various types of helicities to topology 
is a source of non-renormalization theorems.
In particular, the anomalous term in the charge (\ref{operator})
in case of magnetostatics can be rewritten (by using (\ref{deltaL})) 
as a 3d topological photon mass, see, 
in particular, \cite{khaidukov} and references therein.
Furthermore, consider currents of the form 
$J_i(x)~=~I\int d\tau\delta^3(\vec{x}-\vec{x}(\tau)\dot{x}_i(\tau)$.
Then the interaction term of two current loops is given by:
 \begin{eqnarray}
\label{linking}
V=\frac{2II'}{\sigma_M} \int_{C} \!\!\int_{C'} dx^i dy^j 
\epsilon_{ijk} \frac{(x-y)^{k}}{4 \pi |x-y|^3}~,
\end{eqnarray}
where $\sigma_M$ is defined in (\ref{cme}).
The integral in (\ref{linking}) is apparently proportional to the
Gauss linking number of the two current circuits. Moreover
one can demonstrate
\cite{khaidukov}
that the interaction term (\ref{linking}) is not renormalized to any order
in electromagnetic interactions.

To summarize this section, consideration of the chiral anomaly
in the hydrodynamic limit
led us to include into the definition of the conserved axial charge
fluid, magnetic and mixed helicities. All three types of helicities
are conserved in the zero-temperature limit of perfect magnetohydrodynamics.
 It is amusing that
the chiral anomaly unifies all types of helicities which were considered
separately so far. 

It is tempting to reverse the logic and assume that the chiral anomaly
in the hydrodynamic approximation mixes up charges,
associated with helical motions,  which are separately
conserved classically. It is known that $Q^A_{naive}$ is 
indeed conserved classically. 
As is discussed above, other contributions to the axial charge (\ref{extended}) are
conserved in the absence of dissipation. 
Therefore, following this logic one would predict, in particular, that for chiral liquids
\begin{equation}\label{ratio}
\Big(\frac{\eta}{s}\Big)_{classically}~\to~0~
\end{equation}
Such a solution has an advantage of naturally incorporating dissipation-free
chiral magnetic current (\ref{cme}). Note that Eq. (\ref{ratio})
is in no contradiction with the famous lower bound \cite{kovtun} on the 
same ratio $\eta/s$. Indeed, one invokes the quantum-mechanical
uncertainty principle to establish existence of a lower bound on $\eta/s$
\cite{kovtun}.

\section{Instabilities of chiral plasma}

As is mentioned in the Introduction,
chiral plasma can be unstable if one starts with a state where, say,
$Q^A_{naive}\neq 0$ while all other terms in the expression (\ref{extended})
are vanishing. This kind of instabilities has been discussed
recently \cite{shaposhnikov,akamatsu,khaidukov}, see also \cite{redlich}.

We have a few  points to add:
\begin{itemize}
\item{The state with $Q^A_{naive}\neq 0, Q^A_{fh}=Q^A_{mh}=Q^A_{fmh}=0$ can
 decay not only into
the domains with non-vanishing magnetic field
\cite{shaposhnikov,akamatsu,khaidukov}
but also into domains with helical motion
of the plasma, so that $Q^A_{fh}\neq 0$.  }
\item{In particular, we expect that not only primordial magnetic field could 
be produced from
an original right-left asymmetric state \cite{shaposhnikov},
 but primordial helical motion could be generated
on a cosmological scale as well.}
\item{It is amusing to observe, again, that transitions among certain kinds of 
helicities have
been discussed in the literature, independent of the issue of chiral media. 
Starting from conservation of the extended axial charge (\ref{extended})
allows to introduced all possible instabilities in a systematic way.
In particular,
in paper \cite {zeldovich} there is a 
rather detailed discussion  of generation of magnetic field from
the initial helical motion. In our language, this is about the instability:
$$Q^A_{fh}~\to~Q^A_{mh}$$}
\item{The  novel point brought by 
the consideration of chiral media above 
is the transition of $Q_{naive}\neq 0$ to
other components of the conserved axial charge (\ref{extended}).}
\end{itemize} 
The most interesting novel example is the transition of the axial
charge of elementary constituents into right- (or left-) handed vortices:
\begin{equation}\label{vortical}
Q_{naive}~\to~Q_{fh}  .
\end{equation}
Superficially, this transition is similar to  the decay of the axial charge
of the constituents to decay into the magnetic heleicity, $Q_{naive}~\to~Q_{mh}$
discussed above. On the dynamical level, however, there is an important difference.
In case of the electromagnetic instability, $Q_{naive}~\to~Q_{mh}$, one can 
find the unstable mode explicitly, see \cite{akamatsu} and references therein.

In case of the vortical instability (\ref{vortical}) there is no analytic
expression for the unstable mode.  The reason is that there is no perturbation theory
for vortical modes because of the infrared divergences. This is known since long, for
a recent exposition see, e.g., \cite{endlich} and references therein. Roughly speaking, if
in field theory one starts with (an infinite number of) oscillators, 
in case of hydrodynamics one starts with (an infinite number of) free particles. As a result,
dynamics is decided by non-linearities and no analytical methods exist.
Thus, one would turn to numerical methods. Numerical estimates seem especially crucial
since we are considering a rather unusual transformation of motion 
of elementary constituents
into the motion of macroscopic vortices. One could  
suspect that the decay rate of the 
axial charge accumulated in the constituents into vortices 
is very low.

Recently, an important progress was reached in answering this type of question,
see \cite{burch}. Namely, one starts numerical simulations
from a medium of sound waves at a certain temperature. In other words, it is
only microscopic degrees of freedom that are excited first ordinary
. It is found that
vortices are emerging as a result of interactions.
A crucial point is that one starts with the effective Lagrangian derived from first principles, 
see, e.g., \cite{endlich}.  
Moreover, there is a general observation on equivalence of an
irrotational heated system, with no
chemical potential, and (correspondingly adjusted) chemical potential at zero temperature
\cite{rattazzi}. 
Thus, we can say that there is evidence that a system with a non-zero 
chemical potential is unstable with
respect to generation of macroscopic vortices.

This is  not yet the instability we predict.
But it shares the basic dynamical feature, namely,
 transformation of a microscopically chaotic
motion into a macroscopically organized motion.  
As a result of the instability seemingly observed in Ref. \cite{burch}, an equal number of
left- and right-handed vortices is produced
since the total axial charge of the system is zero. What we predict, is that if one starts
with a non-zero axial chemical potential, vortices are produced which are preferably left-
(or right-, depending on the sign of
the axial potential) handed.

\section{Conclusions.}

We a argued that chiral irrotational liquids are unstable
with respect to spontaneous generation of vortices.
Analytically, however, it is not possible to estimate 
the decay rate since it involves transformation of motion of microscopical
degrees of freedom into a macroscopic motion of vortices.
Very recent
lattice simulations
\cite{burch}   indicate
that if one starts with a state of perfect fluid with 
sound waves at a certain temperature and no vortices, vortices
are generated dynamically. Thus, the transformation of a microscopic 
chaotic motion
into an organized macroscopic motion is not suppressed in this case. 
Because of the similarity of underlying mechanisms
  the spontaneous generation of non-vanishing 
macroscopic fluid helicity
from chiral liquids with $\mu_5\neq 0$ would 
seemingly be  not suppressed either.

\section{Acknowledgments}
The authors are grateful to G.E.~Volovik for useful discussions. 
The work on this paper has been partly supported by 
RFBR grant 14-02-01185 and RFBR grant 14-01-31353-mol\_a.

\end{document}